\documentstyle[prabib,aps]{revtex}

\def\vec#1{{\rm\bf #1}}
\begin{document}

\title{Some properties of membranes in nematic solvents}
\author{Peter D.~Olmsted  and  Eugene M.~Terentjev \\
     {\small Cavendish Laboratory, University of
Cambridge, Madingley Road, Cambridge, CB3 0HE, U.K.}}
\maketitle

\begin{abstract}
The fluctuation spectrum of membranes in nematic
solvents is altered by the boundary condition imposed on
the bulk nematic director by the curved membrane. We
discuss some properties of single and multi-membrane
systems in nematic solvents, primarily based on the
Berreman-de~Gennes  model. We show that: membranes in
nematic  solvents are more rigid and less rough than in
their isotropic counterparts; have a  different Helfrich
steric stabilization energy, proportional to $d^{-3}$,
and hence a different compression modulus in the lamellar
state; and can exhibit phase separation via unbinding
during a quench into the nematic state. We also discuss
the preparation and  possible experimental effects of
nematic-mediated surfactant membrane system.
\end{abstract}


\vskip 5truemm
\noindent Short Title: \ \ {\bf Membranes in nematic
solvents}

\newpage
There are many instances in physics where the
configuration of  a field on a boundary is influenced by
the fluctuations of a conjugate field in    the adjacent
volume. Examples of this are the Casimir effect, where
the confinement of the electromagnetic field between
conducting plates results in a net attraction
\cite{casimir}: related effects in soft condensed matter
include  a fluctuation-enhanced interaction between
inclusions on a membrane  \cite{goulian93} and the the
interaction between surfaces dipped in a structured fluid
due to the change in fluctuation spectrum of the fluid
\cite{ajdari92,likardar}. While these are {\sl
entropic\/} in origin, there is another class of
fluctuation enhancements due to the {\sl energy\/} of
deforming the bulk field coupled to the boundary field by
some anchoring condition. Examples here include:
interaction between membrane inclusions due to the strain
induced in the membrane \cite{nilymembranes}; the
non-analytic contribution to the wetting contact line
elasticity due to deformations of  the adjacent fluid-air
interface \cite{joannydegennes84}; and the  example which
we explore here, a non-analytic contribution to the free
energy of a surface in contact with a liquid crystalline
solvent \cite{pgdg,berreman72}.

In this article we explore some of the properties of
fluid membranes in contact with nematic liquid
crystalline solvents, discussing both entropic and
energetic effects. By `membrane' we envision
surfactant bilayers arranged in the archetypical
structures found in surfactant systems: in this work we
focus on lamellar phases. It is well known that
surfactants induce varying degrees and strengths of
boundary conditions on the nematic director
\cite{oleg83}.  Here we consider the simple natural case
when the mesogenic molecules are strongly anchored by the
hydrophobic tails of surfactant in the direction   along
the membrane normal.

Our starting point is the well-known Berreman-de~Gennes
model \cite{pgdg,berreman72}, which was
introduced to describe the anchoring energy of a liquid
crystal on grooved substrates. Here we consider the
`grooved substrate' to be thermal undulations of a
bilayer surface, and hence consider some of the
consequences of the deformation of the nematic director
field, induced by these undulations,  Fig.1. The
principle is this: the equilibrium thermodynamics of a
membrane--liquid-crystalline-solvent system includes as
fluctuating variables both the solvent and the membrane.
If we are interested in properties of the membrane we can
`integrate out' the solvent degrees of freedom to find an
effective theory for the membrane thermodynamics. To
perform such an integration and speak of a renormalized
theory of membranes requires a separation of timescales.
That is, if we are interested in dynamical properties of
the membrane, we must ensure that the solvent
fluctuations (in this case, the establishment of a
deformed director field in response to a surface
undulation) are much faster than the characteristic
membrane decay time. However, if we are interested in
{\sl equilibrium \/} effects, such as fluctuation spectra
as would be measured in experiments lasting `long' times,
then this procedure is valid. It is with these kinds of
experiments in mind that we proceed. Let us first recall
the fundamental ideas of the Berreman-de~Gennes model.

\underline{\sl Nematic Energy\/}---Consider a surface
with a modulation of wavevector $q_{\perp}$, in contact
with a  nematic solvent, and assume strong homeotropic
boundary conditions,
$\delta\bbox{\hat{n}}(\vec{r}_{\perp},z=0)=
-\bbox{\nabla}_{\perp} u(\vec{r}_{\perp}) \,$, where
$\delta\bbox{\hat{n}}$ is a variation of the nematic
director and $u(\vec{r}_{\perp})$ is a surface
displacement along its equilibrium normal $\hat{z}$; the
dimensions in the membrane plane are denoted  by
$\vec{r}_{\perp}$. The bulk nematic solvent minimizes
the Frank elastic energy $F_{\mit F}=(1/2)
K_{\scriptscriptstyle F}\int\! d^3\!r
(\nabla\bbox{\hat{n}})^2 \,$, \cite{pgdg}, in which we
make the one constant approximation for the Frank
constants $ K_{\scriptscriptstyle F}$. The solution is
\begin{equation}
\delta\bbox{\hat{n}}(\vec{r}_{\perp},z)
= \int_{q_{\perp}}\! i\vec{q}_{\perp}  u(q_{\perp})
e^{-i\vec{q}_{\perp}\cdot\vec{r}_{\perp} - |q_{\perp}|z},
\end{equation}
where $\int_{q_{\perp}}\equiv\int\!
d^2\!q_{\perp}/(2\pi)^2$  with the limits between an
upper cutoff  $2\pi/a$ and a lower cutoff
$2\pi/L_{\perp}$, with $a$ being a microscopic dimension
and $L_{\perp}$ the membrane size. Substituting  into the
Frank energy and integrating over the dimension $z$
normal to the  interface, we find
\begin{equation}
F_{\mit F} = {1\over 2} K_{\scriptscriptstyle F}
\int_{q_{\perp}} |q_{\perp}|^3 (1 - e^{-|q_{\perp}| L_z})
|u(q_{\perp})|^2  \ \approx {1\over 2}
K_{\scriptscriptstyle F} \int_{q_{\perp}}  {q_{\perp}^4
L_z\over 1 + |q_{\perp}| L_z}  |u(q_{\perp})|^2 ,
\label{eq:Feff1}
\end{equation}
where $L_z$ is a large
distance cutoff which we take below to be $L_z=\infty$
for an isolated membrane or $L_z=d$ for a stack of
membranes spaced by $d$.  Eq.~(\ref{eq:Feff1})
interpolates between the $\sim |q_{\perp}|^3 $ regime
for an isolated membrane and the $\sim |q_{\perp}|^4L_z$
long wavelength behavior in a finite system. The second
expression in (\ref{eq:Feff1}) is an  alternative
approximation  which handles properly the large and
small~$q$ limits,  and is much easier for calculations.
We shall mostly use this form of Eq.~(\ref{eq:Feff1}) in
this work, since we are primarily concerned with
qualitative results. To this must be added the Helfrich
energy of the fluctuating membrane,
\begin{equation}
F_{\scriptscriptstyle H}={1\over 2}\kappa\int\!
d^2\!r_{\perp} (\nabla_{\perp}^2 u)^2 + \int
d^2\!r_{\perp}\,\bar{\kappa}\,G, \label{eq:Fhelfrich}
\end{equation}
where the Gaussian curvature $G$
integrates to zero for lamellar systems without
topological defects and  plays no further role here.

The unusual non-analytic form of Eq.~(\ref{eq:Feff1})
arises from  the same considerations as the linear-$|q|$
elasticity of the air-fluid-solid triple line, where
energy is stored in the deformation of the air-fluid
surface \cite{joannydegennes84}.  From this energy one
finds many respects in which membranes in nematic
solvents differ from their isotropic counterparts. This
behavior could most easily be seen by preparing  mixtures
of surfactant and thermotropic liquid crystal, with the
latter playing the role of an oil (possibly with water
or a cosurfactant, as is  common in conventional
surfactant systems, to select from the zoo of possible
phases), and cycling  through the solvent's
isotropic-nematic transition temperature. We proceed by
briefly describing some of these properties in order of
complexity. We only consider the case of membranes
without surface tension, where this new term is most
important. In  the Appendix we present the contribution
due to the Casimir effect \cite{likardar}, which is an
entropic effect in the correlated fluid mediating the
membrane, and show that the main effects of it are a
renormalization of the area per surfactant head group,
and the bending modulus $\kappa$.

\underline{\sl Single-membrane properties\/}---For these
properties we take $L_z=\infty$. The first obvious new
effect is a qualitative change in the surface
fluctuations. Bilayers are typically rough due to thermal
fluctuations, and the combination of a two-dimensional
surface fluctuating in three dimensions  yields divergent
height fluctuations. In a nematic solvent, however, this
changes. For example, fluctuations of the surface normal
are given by
\begin{eqnarray}
\langle|\delta\bbox{\hat{n}}(\vec{r}) - \delta\bbox{
\hat{n}}(0)|^2\rangle = 2k_{\scriptscriptstyle B}T
\int_{q_{\perp}} {1-\cos\vec{q}_{\perp}\cdot\vec{r}\over
K_{\scriptscriptstyle F}|q_{\perp}| + \kappa q_{\perp}^2}
\ \
 \simeq   {k_{\scriptscriptstyle B}T \over \pi \kappa}
\log\left[ {K_{\scriptscriptstyle F} + 2\pi\kappa/a
\over K_{\scriptscriptstyle F} +  2\pi\kappa/r}\right],
\end{eqnarray}
where $a$ is a microscopic cutoff. If we define the
correlation length $\xi_0$ as that distance along the
membrane for which fluctuations in the normal vector
$\vec{n}$ orientation are of order $1$
\cite{degennestaupin82},  we find
\begin{equation}
{\xi_0\over a} = {e^{2\pi\kappa/(k_{\scriptscriptstyle
B}T)}\over 1 - {K_{\scriptscriptstyle F} a\over
2\pi\kappa}( e^{2\pi\kappa/(k_{\scriptscriptstyle
B}T)}-1)}.     \label{XiLength}
\end{equation}
For $K_{\scriptscriptstyle F}=0$ the membrane in the
isotropic non-correlated solvent is crumpled at distances
larger than $\xi_0$. For $K_{\scriptscriptstyle F}\neq
0$, in a nematic solvent, $\xi_0$ increases rapidly and
reaches the system size ($\infty$) for
$K_{\scriptscriptstyle F}a/(2\pi\kappa)=
(e^{2\pi\kappa/k_{\scriptscriptstyle B}T} -1)^{-1}$.
Hence, for  $K_{\scriptscriptstyle F}a/\kappa\gg 1$, in
the regime that should be identified with a `strong
nematic solvent', the membrane would not be  crumpled at
all.  A typical estimate of the surfactant bilayer
bending rigidity is $\kappa \sim 5 \cdot 10^{-20} \,
\hbox{J}$, only slightly larger than the thermal energy
at room temperature. Taking a characteristic value for
the Frank constant,  $K_{\scriptscriptstyle F} \sim
10^{-11} \, \hbox{J/m}$ and the molecular size $a \sim 10
\hbox{\AA}$, one obtains an estimate of order unity and,
therefore, both crumpled and flat regimes are accessible
for a membrane in  a nematic solvent.

Also of interest is the related quantity, the membrane
roughness, given by the mean-square height fluctuations:
\begin{eqnarray}
\langle u(\vec{r})^2 \rangle  \
&=& k_{\scriptscriptstyle B}T \int_{q_{\perp}} {1\over
\kappa q_{\perp}^4 +  K_{\scriptscriptstyle F}
|q_{\perp}|^3}   \nonumber  \\
&=& {k_{\scriptscriptstyle
B}T \over 2\pi K_{\scriptscriptstyle F}^2} \left\{
K_{\scriptscriptstyle F} {L_{\perp}\over 2\pi} + \kappa
\log\left[{2\pi\kappa  + a K_{\scriptscriptstyle F} \over
2\pi\kappa + L_{\perp} K_{\scriptscriptstyle F}}\right]
\right\},
\end{eqnarray}
where $L_{\perp}$ is the transverse membrane dimension,
coming from the lower cutoff in $q$-space. As
$K_{\scriptscriptstyle F}\rightarrow 0$ we recover, after
expansion in powers of $L_{\perp} K_{\scriptscriptstyle
F}$, the result for a conventional membrane, $\langle
u^2\rangle \sim L_{\perp}^{2\zeta_S}$, with a roughness
exponent $\zeta_S=1$. In a strong nematic solvent we have
$\zeta_S=1/2$ and, as  expected, the membrane is not as
rough.

\underline{\sl Renormalization of bending
modulus\/}---Since the membrane in the nematic solvent is
stiffer, we expect the renormalization of $\kappa$ due to
thermal  fluctuations \cite{helfrich85} to be much
reduced. There are two new sources of renormalization for
membranes in nematic solvents: entropic, due to the
Casimir effect, which we briefly discuss in the appendix;
and energetic, due to the Berreman-de~Gennes energy.
Following the simple procedure outlined by Helfrich
\cite{helfrich85}, we find
\begin{equation}
\kappa_{\scriptscriptstyle R}=\kappa -
{k_{\scriptscriptstyle B}T\over 4\pi}\left[
\log \left({\kappa q_{max} + K_{\scriptscriptstyle F} \over
\kappa q_{min} + K_{\scriptscriptstyle F}}\right) - {3\over
32}\log{L_{\perp}\over a} \right] ,  \label{renormK}
\end{equation}
where the
first correction is from the Helfrich renormalization and
the second term is produced by the Casimir effect in the
correlated solvent.
In the limit $K_{\scriptscriptstyle F}L_{\perp}/\kappa
\ll 1$ (isotropic solvent) the Helfrich effect returns to the
usual $\log L_{\perp}/a$ reduction of the bending rigidity
\cite{helfrich85,peliti}. In the nematic solvent with
$K_{\scriptscriptstyle F} \neq 0$ it is replaced by the constant
factor $ \log \big[ 1 + (\kappa/K_{\scriptscriptstyle F} a\big] $, so
that the renormalization $\kappa_{\scriptscriptstyle R} - \kappa$
can be large or small depending on the `strength'
of nematic solvent. In addition, there is an {\sl
increase} of $\kappa$ due to the Casimir effect, also
logarithmically divergent with the system size. This
result supports the intuitive expectation for the
membrane to become more rigid due to the anchoring with
the nematic solvent.

\underline{\sl Lamellar Phase:~Helfrich
Interaction\/}---In a lamellar state we take $L_z=d$  as
the cutoff in Eq.~(\ref{eq:Feff1}), since the range  of
the solvent extends only up to neighboring membranes. We
define a `strong'  nematic solvent in this context as one
for which $K_{\scriptscriptstyle F}d\gg\kappa$ [which is,
in fact, a much weaker  condition than
$K_{\scriptscriptstyle F}a\gg\kappa$ in
Eq.(\ref{XiLength})].  To estimate this we again take
$\kappa\sim10^{-20}\,\hbox{J}\, , K_{\scriptscriptstyle
F}\sim 10^{-11} \,\hbox{J/m}$, and lamellar spacings
ranging from $10-1000\, \hbox{\AA}$, yielding
$K_{\scriptscriptstyle F}d/\kappa\sim 1-100$. Since the
moduli $\kappa$ can be changed by adding co-surfactant
and Frank constants depend on the nematic order
parameter, it is quite easy to span the whole range from
weak to strong nematic solvents.

In stacked lamellar
phases there are two well-known interactions which
stabilize the lamellar phase: electrostatic stabilization
and steric interaction. The electrostatic effect
\cite{parsegian79} yields an interaction energy per unit
area of $F/A\sim  k_{\scriptscriptstyle B}T
/(L_{\scriptscriptstyle B} d)$, where
$L_{\scriptscriptstyle B}$ is the Bjerrum length and $d$
is the membrane  separation. We shall not consider this
interaction, exploring instead  the more interesting
statistical effects of electrostatically screened
membranes \cite{roux91}. In an isotropic solvent these
effects lead to the steric stabilization
\cite{helfrich78}, giving $F/A\sim
(k_{\scriptscriptstyle B}T)^2/(\kappa d^2)$. This steric
interaction arises due to the divergence of the height
fluctuations of a single membrane, $\langle
|u|^2\rangle\sim L_{\perp}^{2\zeta_S}$; membranes in a
stack  experience collisions with a characteristic length
between collisions governed  by $\zeta_S$. Since
$\zeta_S$ is positive we expect  a steric interaction in
the nematic solvent as well, but with a different
character than for the standard Helfrich interaction.

Rather than going through a calculation similar to
Helfrich's, we content  ourselves here with a
scaling-type analysis to obtain the $d$-dependence  of
the steric stabilization of membranes in a nematic
solvent.  We first  calculate the height fluctuations,
using the equipartition theorem for the membrane
Hamiltonian, given by Eqs.~(\ref{eq:Feff1}) with $L_z=d$
and~(\ref{eq:Fhelfrich}), and integrating over wave
vectors $\vec{q}_{\perp}$
\begin{equation} \langle
u(\vec{r})^2\rangle \approx {k_{\scriptscriptstyle
B}T\over 4 \pi \kappa q_0^2}
 {1 + [ K_{\scriptscriptstyle F}d /\kappa ] (1 + 2 q_0 d) \over (1 +
[K_{\scriptscriptstyle F}d/\kappa ])^2 },
\end{equation}
where $q_0 = 2\pi/L_{\perp}$ is the low-$q$ cutoff in
the membrane plane (we have ignored subdominant terms
logarithmic in $q_0$). Now, the membrane will be
sterically  stabilized when the height fluctuations are
of order the layer spacing,  $\langle u^2\rangle \simeq
d^2$. This determines $q_0$. For a weak nematic solvent,
$K_{\scriptscriptstyle F} d\ll \kappa$, we recover  the
Helfrich result $q_0 d \approx
\sqrt{k_{\scriptscriptstyle B}T /  \kappa}$. For a strong
nematic solvent we find
\begin{equation} (q_0 d)^2 =
{k_{\scriptscriptstyle B}T \over 4 \pi
K_{\scriptscriptstyle F} d}  (1 + 2 q_0 d)\, ;  \qquad
K_{\scriptscriptstyle F} d \gg\kappa \, .
\end{equation}
Since, typically,  $\kappa >            
k_{\scriptscriptstyle B}T$ we have $
K_{\scriptscriptstyle F} d \gg k_{\scriptscriptstyle B}T
$, in which case  $q_0 \approx  \left(
k_{\scriptscriptstyle B}T / 4 \pi K_{\scriptscriptstyle F
} \right)^{1/2} d^{-3/2}$. This defines a new length $\xi
= q_0^{-1}$, the in-plane correlation length for height
fluctuations, or mean distance between collisions. Now we
may compute the interaction energy. Crudely, the pressure
$P$ due to undulations may be calculated as that of a gas
of sterically interacting  `discs' of dimension $\xi$.
This yields $P \sim k_{\scriptscriptstyle B}T  / (d\,
\xi^2)$. Using our result for $\xi$, and realizing that
the free  energy per unit area is then given by $F/A = P
d$, we arrive at
\begin{equation} {F\over A} = {
(k_{\scriptscriptstyle B}T)^2 \over 4 \pi
K_{\scriptscriptstyle F} d^3}, \label{eq:newHelf}
\end{equation}
which should be compared with the $d^{-2}$
and $d^{-1}$ behavior of, respectively, the standard
Helfrich and electrostatic stabilizations.

The conclusion is that, for electrostatically screened
membranes in a nematic solvent, much closer lamellar
packing can be achieved, which again confirms the
intuitive expectation for more rigid and flat membranes.
The resulting ``smectic'' lamellar phase will have a
different layer compression modulus of the corresponding
Landau-Peierls elastic energy
\begin{equation} F_{\mit
sm}={1\over 2}\int\! d^3\!q \left[ \bar{B} q_z^2  + (K +
K_{\scriptscriptstyle F}) q_{\perp}^4 \right]
|u(q)|^2,    \label{Land-P}
\end{equation}
where $K=\kappa/d$ is the layer bending modulus in the
absence of nematic solvent. In writing this we have
ignored the $|q_{\perp}|d$ term in the denominator of
Eq.~(\ref{eq:Feff1}), since smectic elasticity is
concerned with wavelengths much larger than the smectic
spacing, $|q_{\perp}|d\ll 1$. The compression modulus is
$\bar{B}=B-C_c^2\chi$, where $B$ is the bare compression
modulus, and $\bar{B}$ includes the renormalization due
to the coupling between solvent composition and layer
spacing \cite{nallet90}. $\bar{B}$ is given essentially
by the pressure of the gas of colliding membranes, which
scales as  $\, \bar{B}\sim 1/d^{\rho} \, $, where
$\rho=2, 3, 4$ for electrostatic, standard Helfrich, and
nematic-solvent membranes Eq.(\ref{eq:newHelf}). The
compression modulus may be measured by, for example,
small angle scattering \cite{roux91}.

More useful information can be extracted from the line
shape of the diffusion scattering peak, following from
the Landau-Peierls energy (\ref{Land-P}). As in ordinary
smectics, the structure factor behaves as, for example,
$S(0,q_z)\sim   (q_z - q_d)^{-2+\eta} \,$ with the usual
Caille exponent \cite{caille},
\begin{equation}
\eta = q_d^2 \frac{k_{\scriptscriptstyle B}T }{
8\pi \left[ \bar{B} (K+K_{\scriptscriptstyle F})
\right]^{1/2} } \ .
\end{equation}
Because $K_{\scriptscriptstyle F}\gg K$ in strong
nematic solvents,  $\eta$ should strongly decrease when
the solvent undergoes a nematic transition, leading to a
more rapid decay of the structure factor, as noted above.
Notice also that the exponent depends on $\bar{B}$, which
changes its qualitative dependence on $d$ as one moves
into the strong solvent regime. By measuring $\eta$ and
the penetration depth
$\lambda=\sqrt{\bar{B}/(K+K_{\scriptscriptstyle F})}$
(which may be extracted  by the corrections to the
low-$q$ behavior of $S(0,q_z)$ \cite{rouxsafinya88}) one
may determine both $\bar{B}$ and $K+K_f$ by systematic
dilution and temperature experiments.

\underline{\sl Unbinding Transition\/}---Upon lowering
the temperature into the nematic  phase, the steric
repulsion energy $F/A$ at fixed $d$ drops by a factor
$f_{\scriptscriptstyle N}/f_{\scriptscriptstyle I} \sim
\kappa/(d  K_{\scriptscriptstyle F})\ll 1$. This dramatic
decrease should affect the unbinding of layers
\cite{milnerroux92,lipowskyleibler86}.  Within a
Flory-type theory, Milner and Roux showed that one can
write  the free energy per volume $f$ of a stack of
bilayers as \cite{milnerroux92}
\begin{equation}
f =
{F\over d A} = c {(k_{\scriptscriptstyle
B}T)^2\over\kappa\delta^3} \phi^3 - \chi
k_{\scriptscriptstyle B}T\phi^2,
\end{equation}
where $c$
is a numerical constant, $\delta$ is the bilayer
thickness, and $\chi$ accounts for contributions to the
second virial coefficient  from other than steric ({\it
i.e.\/} typically van~der~Waals) interactions. Here
$\phi\simeq \delta/d$ is the surfactant volume fraction,
and the $\phi^3$ term follows from the isotropic-solvent
steric interaction. When the solvent  undergoes a
transition into the nematic phase the $\phi^3$ term
should  be replaced by \begin{equation} c'
{k_{\scriptscriptstyle B}T\over K_{\scriptscriptstyle
F}\delta^4} \phi^4, \end{equation} where $c'$ is another
numerical constant. Since $K_{\scriptscriptstyle F}d$
can be much larger than $\kappa$, the result is a smaller
repulsion and a smaller preferred interlayer spacing.

At fixed $\phi$ the characteristics of the unbinding
transition are: at low $\chi$ a single bound phase
exists, while $\chi_c=3 c k_{\scriptscriptstyle B}T \phi/
(\kappa\delta^3)$ marks the spinodal line at which the
system phase separates into  bound $(\phi\neq 0)$ and
unbound $(\phi=0)$ phases.

For nematic solvents the spinodal line is given by
\begin{equation}
\chi_c = {6 c' k_{\scriptscriptstyle B}T \over
K_{\scriptscriptstyle F}\delta^4} \phi^2,
\end{equation}
which allows for the possibility of, for example,
phase separation by quenching the solvent into a nematic
state.  The resulting dynamics would be very complicated,
due to the simultaneous nematic coarsening and phase
separation.

\underline{\sl Conclusions}---In summary, a membrane in
a nematic solvent should be much stiffer than in an
isotropic solvent, leading to its different scaling
behavior. Layered systems are in this stiff  regime when
$K_{\scriptscriptstyle F}d/\kappa\gg 1$, which should be
experimentally realizable. The stiffening can be seen in
several quantities,  such as the correlation length
$\xi_0$ for surface normal fluctuations; the roughness
exponent $\zeta_S$; and intermembrane interactions, in
which the standard Helfrich interaction changes its
dependence on the intermembrane spacing $d$. In addition
to the effect of the bulk nematic elastic energy, a
membrane is also affected by the entropic Casimir-like
effect of fluctuations of the coupled director field.  We
have not yet considered the case of smectic solvents, but
it is straightforward to show, by arguments very similar
to those used in deriving Eq.~(\ref{eq:Feff1}), that
there is a simple renormalization of the membrane
elasticity modulus, $\kappa_{\scriptscriptstyle R}=
\kappa + 2\sqrt{\kappa_s B_s}d$, where the subscript $s$
refers to the solvent smectic elastic constants; and the
Helfrich interaction is the same as in the isotropic
case, with $\kappa$ replaced by
$\kappa_{\scriptscriptstyle R}$.

Interesting effects are expected in the presence of an
external magnetic field $\vec{H}$, which provides a
``mass'' for the nematic director fluctuations. While the
effect of magnetic field on a membrane is, in principle,
the same as for an isotropic solvent, the anisotropy of
the diamagnetic susceptibility of a bilayer membrane
should be negligibly   small compared to that of a bulk
nematic liquid crystal.   It is straightforward to show
that the application of a magnetic field $H$ along the
layer normal yields a term in the  free energy
Eq.(\ref{eq:Feff1}) (for a single membrane in an infinite
system) proportional to
$q_{\perp}^2\sqrt{q_{\perp}^2+\xi_{\scriptscriptstyle
H}^{-2}}$, where $\xi_{\scriptscriptstyle H}=\left(
K_{\scriptscriptstyle F}/(\chi_a H^2) \right)^{1/2}$ is
the standard magnetic coherence length. This term further
reduces the height fluctuations of the membrane to
$\langle |u|^2\rangle\sim\log L_{\perp}/a$, which leads
to a very weak steric Helfrich repulsion  $F/A \sim
1/d^4$, and suppresses the Landau-Peierls instability in
favor of Bragg peaks at the smectic wavevector.

It seems fairly straightforward to perform experimental
checks on the described system, by mixing a thermotropic
nematic with a small concentration of surfactant,
choosing its hydrophobic part to be closely related to
mesogenic molecules. Addition of a small amount of water
would further stabilize the bilayer membrane structure.
All our arguments suggest that it would be very difficult
to create curved micellar structures in the nematic
solvent (spherical micelles, for example, would have to
create a topological defect in the nematic field around
them, due to the radial director anchoring). Instead, we
expect the formation of rather flat bilayers even at very
low concentrations and dense lamellar and sponge phases
with more surfactant/water, with the morphology driven by
the elastic energy effects in the mediating nematic
solvent.  \medskip

We appreciate useful discussions with M. Warner and the
support and practical advice from the Polymers \&
Colloids group of Cavendish Laboratory, which allowed us
to observe some of the described effects in practice.
This research has been financially supported by
Unilever-PLC (EMT) and the EPSRC (PDO).

\newpage
\noindent {\sc Appendix:} \underline{\sl  Casimir
Effect\/}---In addition to the energy stored in the
director field there is an entropic contribution to  the
membrane free energy due to the analog of the Casimir
effect, calculated  for liquid crystals by Ajdari, {\sl
et al.\/} \cite{ajdari92} and Li~and Kardar
\cite{likardar}. From the results of Li and Kardar, the
entropic contribution to the free energy per unit area of
a membrane fluctuating  above a flat surface a distance
$d$ away is \cite{likardar}
\begin{eqnarray}
{F \over
k_{\scriptscriptstyle B}T A} & = & -{a_1\over d^2}\left[1
+ {3 A\over 2 d^2}\int_{q_{\perp}}
|u(q_{\perp})|^2\right]   \nonumber  \\  && \qquad  +
{A\over 64\pi a^2}\left[1 +  4\pi C_1(1)\left({a\over
d}\right)^2\right]\int_{q_{\perp}}  q_{\perp}^2
|u(q_{\perp})|^2  \nonumber  \\ && \qquad + {3\over
128\pi} \left[ \log{L_{\perp}\over a}  + 4 \pi C_1(2)
\right]  \int_{q_{\perp}}  q_{\perp}^4 |u(q_{\perp})|^2
,\nonumber
\end{eqnarray}
where $a_1=0.04792$, and
$C_1(\zeta)$ is given by Eq.~(2.19) of Ref.~[4b] and,
generally, are very small. To obtain this we have
expanded Li~and Kardar's results, which hold for an
arbitrary surface, in a gradient expansion in the
membrane fluctuation $u(\vec{r}_{\perp})$. The term
multiplied by $a_1$ is essentially the classic Casimir
attraction between the plates, and contributes a small
renormalization to the existing attractive interactions,
which are typically van~der~Waals \cite{israel92}. The
$q_{\perp}^2|u(q_{\perp})|^2$ term  renormalizes  the
surface tension and, since a surfactant system in
solution adjusts its area per head group $\Sigma$ to
retain equilibrium and satisfy vanishing surface tension
\cite{degennestaupin82},  leads only to a slight decrease
in $\Sigma$. The last term, written in the
single-membrane limit $d \rightarrow \infty$,
renormalizes the membrane bending rigidity $\kappa$ by a
relevant logarithmic term (see Eq.~(\ref{renormK}), where
we neglected the small correction $C_1(2) \sim 10^{-3}$),
which should be compared  with the value
$[k_{\scriptscriptstyle B}T / 4\pi]\, \log(L_{\perp}/a)$
found for the renormalization due to thermal fluctuations
of membranes in isotropic solvents \cite{helfrich85}.



\newpage
\noindent \large {\bf FIG. 1} \qquad Director field {\bf
n} near a fluctuating membrane, which imposes homeotropic
anchoring


\begin{thebibliography}{99}
%
\bibitem{casimir}
H.~B.~G. Casimir, {\it Proc. Kon. Ned. Akad. Wet.}  {\bf
51} (1948) 793.
%
\bibitem{goulian93}
M. Goulian, R. Bruinsma, and P. Pincus,  {\it Europhys.
Lett.} {\bf 22} (1993) 145.
%
\bibitem{ajdari92}
A. Ajdari, L. Peliti and J. Prost, {\it Phys. Rev.
Lett.} {\bf 66} (1991) 1481; \ {\it J.~Phys.~II (France)}
{\bf 2} (1992) 487.
%
\bibitem{likardar}
H. Li and M. Kardar, {\it  Phys. Rev. Lett.}  {\bf 67}
(1991) 3275; \
 {\it Phys.~Rev.} {\bf A46} (1992)  6490.
%
\bibitem{nilymembranes}
N. Dan, P. Pincus, and S.~A. Safran, {\it Langmuir}
{\bf 9} (1993) 2768.
%
\bibitem{joannydegennes84}
J.~F. Joanny and P.~G. {de Gennes}, {\it  J.~Chem.
Phys.} {\bf 81} (1984) 552.
%
\bibitem{pgdg}
P.~G. de~Gennes and J. Prost, {\em The Physics of Liquid
Crystals}, 2nd  ed.  (Clarendon, Oxford, 1993).
%
\bibitem{berreman72}
D.~W. Berreman, {\it Phys. Rev. Lett.} {\bf 28} (1972)
1683.
%
\bibitem{oleg83}
G.~E. Volovik and O.~D. Lavrentovich, {\it Sov.~Phys.
JETP} {\bf 58} (1983)  1159.
%
\bibitem{degennestaupin82}
P.~G. {de Gennes} and C. Taupin, {\it J.~Phys. Chem.}
{\bf 86} (1982) 2294.
%
\bibitem{helfrich85}
W. Helfrich, {\it J.~Phys (France)} {\bf 46} (1985) 1263.
%
\bibitem{peliti} L. Peliti and S. Leibler, {\it Phys.
Rev. Lett.}, {\bf 54}, 1690 (1985).
%
\bibitem{parsegian79}
A. Parsegian, N. Fuller, and R.~P. Rand, {\it Proc.
Natl. Acad. Sci.~USA}  {\bf 76}  (1979) 2750.
%
\bibitem{roux91}
D. Roux, {\it  Physica} {\bf A172} (1991) 242.
%
\bibitem{helfrich78}
W. Helfrich, {\it Z.~Naturforsch.} {\bf 33a} (1978) 305.
%
\bibitem{nallet90}
F. Nallet, D. Roux, and S.~T. Milner, {\it J.~Phys
(France)} {\bf 51} (1990) 2333.
%
\bibitem{caille}
A. Caille, {\it C.R. Acad. Sc. Paris} {\bf B274}  (1972)
891.
%
\bibitem{rouxsafinya88}
D. Roux and C.~R. Safinya, {\it J.~Phys (France)}  {\bf
49} (1988) 307.
%
\bibitem{milnerroux92}
S.~T. Milner and D. Roux, {\it  J.~Phys.~I (France)} {\bf
2} (1992) 1741.
%
\bibitem{lipowskyleibler86}
R. Lipowsky and S. Leibler, {\it Phys. Rev. Lett.}  {\bf
56} (1986) 2541.
%
\bibitem{morsemilner94}
D.~C. Morse and S.~T. Milner, {\it Europhys. Lett.}
{\bf 26} (1994) 565.
%
\bibitem{israel92}
J.~N. Israelachvilli and H. Wennerstr{\"o}m,  {\it
J.~Phys. Chem.}  {\bf 96} (1992) 520.
%

\end{thebibliography}
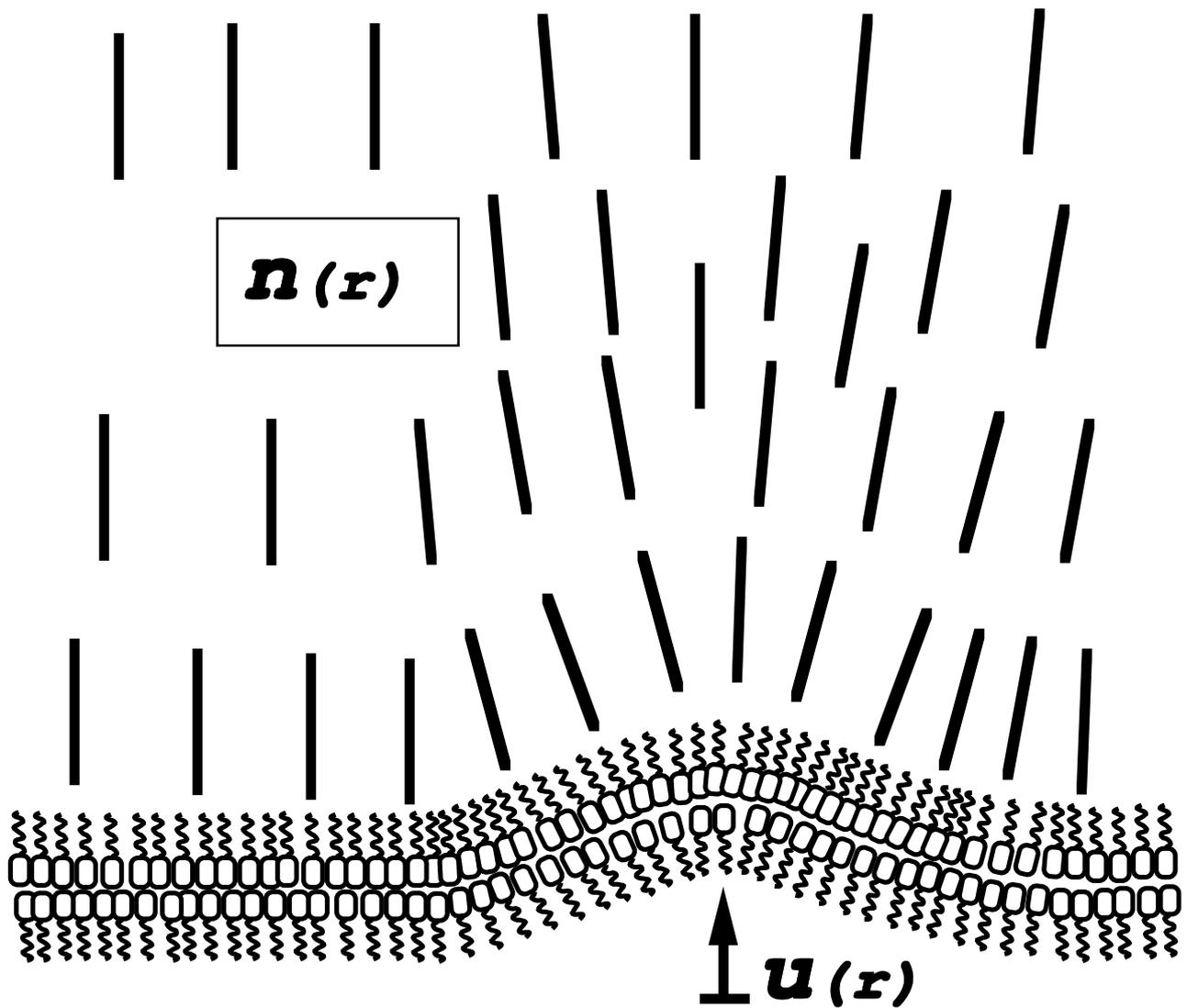
\end{document}